\documentclass[12pt]{iopart}
\usepackage{iopams}
\usepackage{graphicx}
\usepackage{cite}

\newcommand{\re}{\mathop{\mathrm{Re}}}

\begin{document}

\title{$1/f$ noise from point process and time-subordinated Langevin equations}

\author{J Ruseckas, R Kazakevi\v{c}ius and B Kaulakys}

\address{Institute of Theoretical Physics and Astronomy, Vilnius University,
A.~Go\v{s}tauto 12, LT-01108 Vilnius, Lithuania}

\ead{julius.ruseckas@tfai.vu.lt}

\begin{abstract}
Internal mechanism leading to the emergence of the widely occurring $1/f$ noise
still remains an open issue. In this paper we investigate the distinction
between internal time of the system and the physical time as a source of $1/f$
noise. After demonstrating the appearance of $1/f$ noise in the earlier proposed
point process model, we generalize it starting from a stochastic differential
equation which describes a Brownian-like motion in the internal (operational) time.
We consider this equation together with an additional equation relating the
internal time to the external (physical) time. We show that the relation between
the internal time and the physical time that depends on the intensity of the
signal can lead to $1/f$ noise in a wide interval of frequencies. The present
model can be useful for the explanation of the appearance of $1/f$ noise in
different systems.

\end{abstract}

\maketitle

\section{Introduction}

The $1/f$ noise is a random process described by the power spectral density
(PSD), $S(f)$, roughly proportional to the reciprocal frequency, $1/f$, i.e.,
$S(f)\propto1/f^\beta$, with $\beta$ close to $1$. It was observed first as an
excess low-frequency noise in vacuum tubes \cite{Johnson1925,Schottky1926},
later in condensed matter
\cite{Bernamont1934,Bernamont1937a,Bernamont1937b,McWhorter1957,Hooge1981} and
other systems \cite{Weissman1988,Mandelbrot-1999,Scholarpedia2007}. The general
nature of $1/f$ noise (named also ``flicker noise'' and ``$1/f$ fluctuations'')
is up to now the subject of several discussions and investigations, see
\cite{Wong2003,Scholarpedia2007,Kogan2008,Balandin2013} for review.

Many models have been proposed to explain the origin of $1/f$ noise. A short
discussion about the models and theories of $1/f$ noise is available in the
introduction of paper \cite{Kaulakys2009}. Widely used model of $1/f$ noise
interprets the spectrum as a superposition of Lorentzians with a wide range
distribution of relaxation times
\cite{Bernamont1937b,McWhorter1957,Watanabe2005,Kaulakys2005}. Another
possibility to model signals and processes featuring $1/f^\beta$ noise is a
representation of the signals as consisting of the renewal pulses or events with
the power-law distribution of the inter-event time \cite{Lowen2005}.

A class of models of $1/f$ noise relevant for driven nonequilibrium systems
involves the self-organized criticality (SOC)
\cite{Bak1987,Bak1996,Banerjee2006,Huang2015}. SOC refers to the tendency of
nonequilibrium systems driven by slow constant energy input to organize
themselves into a correlated state where all scales are relevant \cite{Bak1996}.
In \cite{Bak1987} a simple driven automaton model of sandpiles that reaches a
state characterized by power-law time and space correlations has been
introduced. However, the mechanism of self-organized criticality not necessarily
results in $1/f^{\beta}$ fluctuations with $\beta$ close to $1$
\cite{Jensen1989,Kertesz1990}. The $1/f$ noise in the fluctuations of a mass was
first seen in a sandpile model with threshold dissipation, proposed in
\cite{Ali1995}. In addition, the exponent $\beta$ is exactly $1$ in the spectrum
of fluctuations of mass in a one-dimensional directed model of sandpiles
\cite{Maslov1999}.

In most cases the $1/f$ noise is a Gaussian process \cite{Kogan2008,Li2012},
however sometimes $1/f$ fluctuations are non-Gaussian
\cite{Orlyanchik2008,Melkonyan2010}. Processes with the power-law distributions
of the signal characteristics can be modeled by presuming that the time between
the adjacent pulses experience slow (the change from one inter-pulse duration to
the next much smaller than the duration itself) Brownian-like motion
\cite{Kaulakys1998,Kaulakys1999,Kaulakys2000-2}. Moreover, the nonlinear
stochastic differential equations (SDEs) generating $1/f^\beta$ noise have been
obtained and analyzed \cite{Kaulakys2004,Kaulakys2006,Kaulakys2009} starting
from this point process model. SDE generating $1/f$ noise should necessarily be
nonlinear, because systems of linear SDEs do not generate signals with $1/f$
spectrum. Such nonlinear SDEs have been applied to describe signals in
socio-economical systems \cite{Gontis2010,Mathiesen2013}.

In the signal consisting of a sequence of pulses the pulse number is a
progressively increasing quantity and it can be understood as an internal time
of the process. The purpose of this paper is to investigate the distinction
between the internal time of the system and the physical time in connection with
$1/f$ noise. We intend to generalize the mechanism leading to $1/f$ noise in the
point process model, proposed in
\cite{Kaulakys1998,Kaulakys1999,Kaulakys2000-2}. Instead of a sequence of pulses
we start from an SDE describing a Brownian-like motion. We compose a new
equation by interpreting the time in the SDE as an internal parameter and adding
an additional equation relating the internal time to the physical time. We
demonstrate that the relation between the internal time and the external time, 
depending on the intensity of the signal, can lead to $1/f$ noise in a wide
interval of frequencies.

A process $x(\tau(t))$ obtained by randomizing the time clock of a random
process $x(t)$ using a new clock $\tau(t)$, where $\tau(t)$ is a random process
with non-negative increments, is called the subordinated process
\cite{Feller1971}. The process $\tau(t)$ is referred to as directing process,
randomized time or operational time. In physics the time-subordinated equations
have been applied to describe anomalous diffusion. Fogedby \cite{Fogedby1994}
introduced a class of coupled Langevin equations consisting of a Langevin
process $x(s)$ in a coordinate $s$ and a L{\'e}vy process representing a
stochastic relation $t(s)$. This class of coupled Langevin equations has been
further investigated in \cite{Baule2005}, where $N$-time joint probability
distributions have been analyzed. Properties of the inverse $\alpha$-stable
subordinator have been considered in \cite{Piryatinska2005,Magdziarz2006}. It
has been shown \cite{Stanislavsky2003, Magdziarz2007} that the description of
anomalous diffusion by a Markovian dynamics governed by an ordinary Langevin
equation but proceeding in an auxiliary, operational time instead of the
physical time is equivalent to a fractional Fokker-Planck equation. Numerical
simulation of subordinated equations has been explored in
\cite{Kleinhans2007,Magdziarz2007}.

In contrast to the description of the anomalous diffusion, in this paper we
consider the situation when small increments of the physical time are
proportional to the increments of the operational time, with the the coefficient
of proportionality that depends on the stochastic variable $x$ representing
the signal intensity. Thus, in our case the randomness of the operational time
comes from the randomness of $x$.

The paper is organized as follows: In \sref{sec:pulses} we briefly present the
point process model of $1/f$ noise and obtain the PSD of the signal by a new
method. In \sref{sec:subord} we generalize the mechanism leading to $1/f$ noise
presented in \sref{sec:pulses}. We introduce the difference between the
physical and the internal time and consider time-subordinated Langevin
equations. In \sref{sec:example} we examine several stochastic processes and,
introducing the internal and external times, we check whether $1/f$ noise can be
obtained. In \sref{sec:numer} we discuss a way of solving highly non-linear SDEs
by introducing suitably chosen internal time and the variable step of
integration. \Sref{sec:concl} summarizes our findings.

\section{1/f noise in a signal consisting of pulses}

\label{sec:pulses}One of the models of $1/f$ noise has been presented in
\cite{Kaulakys1998,Kaulakys1999,Kaulakys2000-2}. In this model a signal consist
of pulses with the time between adjacent pulses undergoing a Brownian-like
motion. It has been shown that this Brownian-like motion of the inter-pulse
durations can yield $1/f$ noise. In this section we briefly present this model
and obtain the PSD of the signal using a different method
than the method used in \cite{Kaulakys1998,Kaulakys1999,Kaulakys2000-2}. The new
method allows us better estimate the frequency range where the PSD has $1/f$
behavior. 

\begin{figure}
\includegraphics[width=0.6\textwidth]{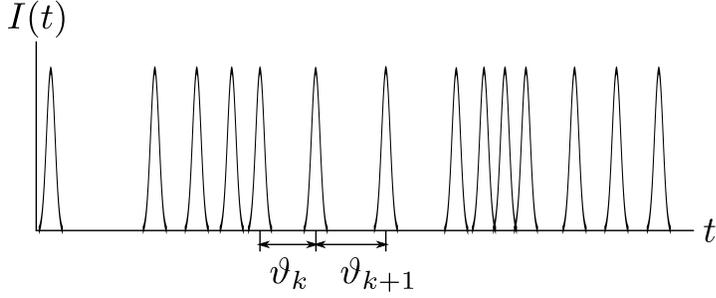}
\caption{Sequence of pulses with random inter-pulse durations $\vartheta_{k}$.}
\label{fig:point}
\end{figure}

Let us consider a signal consisting of a pulse sequence having correlated
inter-pulse durations. We assume that: (i) the pulse sequences are stationary
and ergodic; (ii) all pulses are described by the same shape function $A(t)$.
The general form of the signal can be written as
\begin{equation}
I(t)=\sum_{k}A(t-t_{k})\,,\label{eq:signal}
\end{equation}
where the functions $A(t)$ determine the shape of the individual pulse and time
moments $t_{k}$ determine when the pulse occurs. Inter-pulse duration is
$\vartheta_{k}=t_{k+1}-t_{k}$. Such a pulse sequence is schematically shown in
\fref{fig:point}. 

The PSD of such a signal is given by the equation
\begin{equation}
S(f)=\lim_{T\rightarrow\infty}\left\langle \frac{2}{T}\left|
\int_{t_{i}}^{t_{f}}I(t)\rme^{-\rmi 2\pi ft}\rmd t\right|^{2}\right\rangle \,,
\label{eq:spectr}
\end{equation}
where $T=t_{f}-t_{i}$ is the observation time and the brackets
$\langle\cdot\rangle$ denote the averaging over realizations of the pulse
sequence. Note that in equation \eref{eq:spectr} we consider one-sided PSD, thus
we have multiplier $2$ in it. Introducing the Fourier transform $F(\omega)$ of
the pulse shape function $A(t)$, we can write equation \eref{eq:signal} as 
\begin{equation}
S(f)=|F(\omega)|^{2}\lim_{T\rightarrow\infty}\left\langle \frac{2}{T}
\left|\sum_{k}\rme^{-\rmi\omega t_{k}}\right|^{2}\right\rangle \,.
\end{equation}
Here $\omega=2\pi f$. If the pulses are narrow and we are considering low
frequencies then the Fourier transform $F(\omega)$ of the pulse shape is almost
constant. In this case we can replace the actual pulses with $\delta$-functions
and drop $F(\omega)$ in the equations. 

The PSD can be decomposed into two parts, 
\begin{eqnarray}
S(f) & = \lim_{T\rightarrow\infty}\left\langle \frac{2}{T}
\sum_{k,k'}\rme^{\rmi\omega(t_{k'}-t_{k})}\right\rangle \\
& = \lim_{T\rightarrow\infty}\left\langle \frac{2}{T}\sum_{k}1\right\rangle
+\lim_{T\rightarrow\infty}\left\langle \frac{2}{T}\left(\sum_{k'>k}
\rme^{\rmi\omega(t_{k'}-t_{k})}+\sum_{k>k'}
\rme^{\rmi\omega(t_{k'}-t_{k})}\right)\right\rangle
\label{eq:s-intermed}\\
& \equiv S_{1}(f)+S_{2}(f)\,.
\end{eqnarray}
The first term can be written as
\begin{equation}
S_{1}(f)=2\nu\,,
\label{eq:s1}
\end{equation}
where $\nu$ is the mean number of pulses per unit time. By changing $k$ into
$k'$ in the second part of the PSD one sees that it can be expressed as
\begin{equation}
S_{2}(f)=4\re\lim_{T\rightarrow\infty}\left\langle
\frac{1}{T}\sum_{k'>k}\rme^{\rmi\omega(t_{k'}-t_{k})}\right\rangle\,,
\label{eq:s2}
\end{equation}
where the time difference $t_{k'}-t_{k}$ is
\begin{equation}
t_{k'}-t_{k}=\sum_{q=k}^{k'-1}\vartheta_{q}\,.
\end{equation}
Thus, equation \eref{eq:s-intermed} becomes
\begin{equation}
S(f)=2\nu+4\nu\re\sum_{q=1}^{\infty}\left\langle
\rme^{\rmi\omega\sum_{j=0}^{q-1}\vartheta_{j}}\right\rangle \,.
\label{eq:spectr-inter}
\end{equation}
Assuming that the joint probability
$P(\vartheta_{0},\vartheta_{1},\ldots\vartheta_{q-1})$ exist we can write the
average in the above equation as
\begin{eqnarray}
\fl\left\langle \rme^{\rmi\omega\sum_{j=0}^{q-1}\vartheta_{j}}\right\rangle & = &
\int\rmd\vartheta_{0}\int\rmd\vartheta_{1}\cdots\int\rmd\vartheta_{q-1}
P(\vartheta_{0},\vartheta_{1},\ldots,\vartheta_{q-1})
\rme^{\rmi\omega\sum_{j=0}^{q-1}\vartheta_{j}}\\
& = & \int\rmd\vartheta_{0}
P_{\vartheta}(\vartheta_{0})\rme^{\rmi\omega\vartheta_{0}}
\int\rmd\vartheta_{1}
P(\vartheta_{1}|\vartheta_{0})
\rme^{\rmi\omega\vartheta_{1}}\cdots\nonumber\\
& & \times\int\rmd\vartheta_{q-1}
P(\vartheta_{q-1}|\vartheta_{0},\vartheta_{1},\ldots,\vartheta_{q-2})
\rme^{\rmi\omega\vartheta_{q-1}}\,.
\end{eqnarray}
If the inter-pulse durations follow the Markov process then the conditional
probabilities depend only on the previous value of the inter-pulse duration,
$P(\vartheta_{j}|\vartheta_{0},\vartheta_{1},\ldots,\vartheta_{j-1})=
P(\vartheta_{j}|\vartheta_{j-1})$. In this case
\begin{eqnarray}
\fl\left\langle\rme^{\rmi\omega\sum_{j=0}^{q-1}\vartheta_{j}}\right\rangle =
\int\rmd\vartheta_{0}P_{\vartheta}(\vartheta_{0})\rme^{\rmi\omega\vartheta_{0}}
\int \rmd\vartheta_{1}P(\vartheta_{1}|\vartheta_{0})
\rme^{\rmi\omega\vartheta_{1}}\cdots\nonumber\\
\times\int\rmd\vartheta_{q-1}P(\vartheta_{q-1}|\vartheta_{q-2})
\rme^{\rmi\omega\vartheta_{q-1}}\,.\label{eq:tmp-1}
\end{eqnarray}

Let us consider a situation when the probability density function (PDF) of
inter-pulse durations $P_{\vartheta}(\vartheta)$ is significant only for
$\vartheta$ in some range $\vartheta_{\mathrm{min}}\leq\vartheta\leq
\vartheta_{\mathrm{max}}$ and is very small for $\vartheta$ outside this range.
In addition, we will assume that the conditional probability
$P(\vartheta_{j}|\vartheta_{j-1})$ has the following properties: the average
equal to the previous value of inter-pulse duration 
\begin{equation}
\int P(\vartheta_{j}|\vartheta_{j-1})\vartheta_{j}\rmd\vartheta_{j}=\vartheta_{j-1}
\label{eq:cond-avg}
\end{equation}
and the dispersion
\begin{equation}
\sigma^{2}=\int P(\vartheta_{j}|\vartheta_{j-1})(\vartheta_{j}-\vartheta_{j-1})^{2}
\rmd\vartheta_{j}
\end{equation}
is much smaller than the dispersion of inter-pulse durations 
\begin{equation}
\sigma_{\vartheta}^{2}=\int P_{\vartheta}(\vartheta)(\vartheta-\bar{\vartheta})^{2}
\rmd\vartheta\,.
\end{equation} 
These assumptions denote that the average difference between the neighboring
inter-pulse durations is small, i.e., the increments and decrements of the
inter-event duration are small in comparison to the inter-event time it self. 

When $\vartheta_{\mathrm{max}}\gg\vartheta_{\mathrm{min}}$ then the dispersion
of inter-pulse durations is
$\sigma_{\vartheta}^{2}\sim\vartheta_{\mathrm{max}}^{2}$. Thus, we assume that
$\sigma\ll\vartheta_{\mathrm{max}}$. When the assumptions \eref{eq:cond-avg} and
$\sigma\ll\sigma_{\vartheta}$ hold, we can approximate the conditional
probability $P(\vartheta_{j}|\vartheta_{j-1})$ by a $\delta$-function:
$P(\vartheta_{j}|\vartheta_{j-1})\approx\delta(\vartheta_{j}-\vartheta_{j-1})$.
The approximation in equation \eref{eq:tmp-1} is valid only for sufficient small
$q$, smaller than some maximum value $q_{\mathrm{max}}$, because the error grows
with the number of terms. Using in equation \eref{eq:tmp-1} the approximation of
the conditional probability by $\delta$-function we obtain 
\begin{equation}
\left\langle\rme^{\rmi\omega\sum_{j=0}^{q-1}\vartheta_{j}}\right\rangle \approx
\int_{0}^{\infty}P_{\vartheta}(\vartheta_{0})
\rme^{\rmi\omega q\vartheta_{0}}\rmd\vartheta_{0}=\chi_{\vartheta}(\omega q)\,,
\end{equation}
where
\begin{equation}
\chi_{\vartheta}(\omega)=\int_{0}^{\infty}P_{\vartheta}(\vartheta)
\rme^{\rmi\omega\vartheta}\rmd\vartheta
\end{equation}
is the characteristic function of inter-pulse durations.

We can estimate the value of $q_{\mathrm{max}}$ as follows: the approximation of
the conditional probability $P(\vartheta_{j}|\vartheta_{j-1})$ by
$\delta$-function is not applicable when the dispersion of $\vartheta_{q-1}$ for
a given $\vartheta_{0}$ becomes comparable with the dispersion
$\sigma_{\vartheta}^{2}$. Assuming that the dispersion of $\vartheta_{j}$, for a
given $\vartheta_{0}$, grows linearly with $j$ (as would be the case for a
random walk) we require that
$\sigma^{2}q_{\mathrm{max}}\lesssim\sigma_{\vartheta}^{2}$ and, therefore, 
\begin{equation}
q_{\mathrm{max}}\sim\frac{\vartheta_{\mathrm{max}}^{2}}{\sigma^{2}}\,.
\end{equation}
For high enough frequency, when
\begin{equation}
\omega q_{\mathrm{max}}\vartheta_{\mathrm{max}}\gg1\,,
\end{equation}
the characteristic functions $\chi_{\vartheta}(\omega q)$ corresponding to large
$q\sim q_{\mathrm{max}}$ are small and we can neglect in equation
\eref{eq:spectr-inter} the terms with $q>q_{\mathrm{max}}$. Including only the
terms with $q\leq q_{\mathrm{max}}$ we get the expression for the PSD
\begin{equation}
S(f)\approx2\nu\sum_{q=-q_{\mathrm{max}}}^{q_{\mathrm{max}}}\chi_{\vartheta}(\omega q)\,.
\label{eq:summ}
\end{equation}
After the summation in equation \eref{eq:summ} we obtain
\begin{equation}
\fl S(f)\approx2\nu\int_{0}^{\infty}\frac{\sin\left(\left(\frac{1}{2}+q_{\mathrm{max}}\right)
\omega\vartheta\right)}{\sin\left(\frac{\omega\vartheta}{2}\right)}
P_{\vartheta}(\vartheta)\rmd\vartheta
\approx\frac{4\nu}{\omega}\int_{\omega\vartheta_{\mathrm{min}}}^{\omega\vartheta_{\mathrm{max}}}
\frac{\sin(q_{\mathrm{max}}u)}{u}P_{\vartheta}\left(\frac{u}{\omega}\right)\rmd u\,.
\end{equation}
We have dropped $1/2$ in $\sin(\cdot)$ because $q_{\mathrm{max}}$ is large,
$q_{\mathrm{max}}\gg1$. In addition, for small frequencies
$\omega\vartheta_{\mathrm{max}}\ll1$ we approximated $\sin(u/2)$ in the
denominator as $u/2$. The function $\sin(q_{\mathrm{max}}u)/u$ has a sharp peak
of the width $\pi/q_{\mathrm{max}}$ at $u=0$ and decreases at larger $u$. If
$\omega\vartheta_{\mathrm{max}}\gg\pi/q_{\mathrm{max}}$ then this peak is much
narrower that the width of the PDF $P_{\vartheta}$. In addition, the peak of the
function $\sin(q_{\mathrm{max}}u)/u$ has a significant overlap with
$P_{\vartheta}$ when $\omega\vartheta_{\mathrm{min}}\ll\pi/q_{\mathrm{max}}$. In
this case we obtain the following approximate expression for the PSD:
\begin{equation}
S(f)\approx\frac{4\nu}{\omega}P_{\vartheta}(\vartheta_{\mathrm{min}})
\int_{0}^{\infty}\frac{\sin(q_{\mathrm{max}}u)}{u}\rmd u
=\frac{\nu}{f}P_{\vartheta}(\vartheta_{\mathrm{min}})\,.
\end{equation}
This equation shows that we get $1/f$ spectrum. 

Summing up the assumptions made above, the range of the frequencies where this
expression for PSD holds is
\begin{equation}
\frac{\sigma^{2}}{\vartheta_{\mathrm{max}}^{3}}\ll f\ll
\min\left(\frac{\sigma^{2}}{\vartheta_{\mathrm{min}}\vartheta_{\mathrm{max}}^{2}},
\frac{1}{\vartheta_{\mathrm{max}}}\right)\,.
\label{eq:range}
\end{equation}
When $\vartheta_{\mathrm{min}}<\sigma^{2}/\vartheta_{\mathrm{max}}$ the upper
limit of the frequency range is determined by $\vartheta_{\mathrm{max}}$. In
this case the ratio of upper and lower limiting frequencies is
$\vartheta_{\mathrm{max}}^{2}/\sigma^{2}$. For larger $\vartheta_{\mathrm{min}}$
the ratio of upper and lower limiting frequencies is
$\vartheta_{\mathrm{max}}/\vartheta_{\mathrm{min}}$.

\begin{figure}
\includegraphics[width=0.6\textwidth]{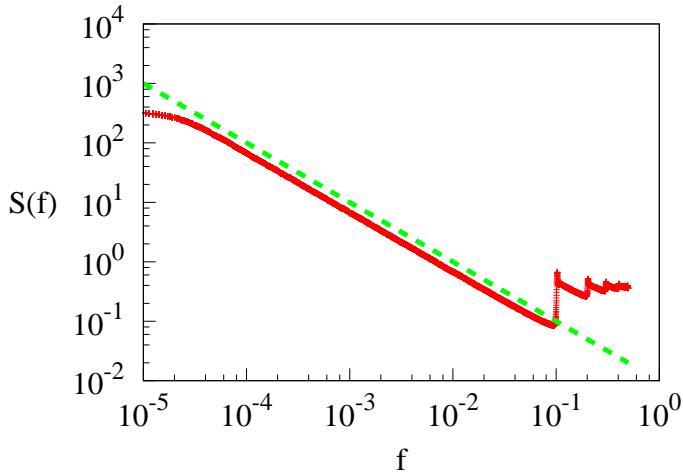}
\caption{The PSD of a signal when the inter-pulse duration performs a random
walk \eref{eq:random-walk}. The dashed (green) line shows $1/f$ spectrum. The
parameters used are $\vartheta_{\mathrm{min}}=0$, $\vartheta_{\mathrm{max}}=10$,
$\sigma=0.1$.}
\label{fig:spectrum-interpulse}
\end{figure}

As an example, let us consider the point process where the inter-pulse durations
perform a random walk and are related via the equation
\begin{equation}
\vartheta_{j+1}=\vartheta_{j}\pm\sigma\,.\label{eq:random-walk}
\end{equation}
Here each sign occurs with probability $1/2$. In addition, we have reflections
from the minimum inter-pulse duration $\vartheta_{\mathrm{min}}=0$ and from the
maximum inter-pulse duration $\vartheta_{\mathrm{max}}$. Numerically obtained
PSD of such a signal is shown in \fref{fig:spectrum-interpulse}. We see a
power-law part in the PSD with the slope $-1$ in a broad range of frequencies
from $4.\times10^{-5}$ to $10^{-1}$. This range of frequencies agrees with the
estimation \eref{eq:range}.

PSD of the power-law form with the exponents different from $-1$ can
be obtained by including in equation \eref{eq:cond-avg} an additional drift term. 
%, that makes the average different from the previous value 
% of inter-pulse duration. 
In \cite{Kaulakys2005} it has been shown that the drift
term of the power-law form $\vartheta^{\delta}$ and power-law PDF of the 
inter-pulse duration $P_{\vartheta}(\vartheta)\sim\vartheta^{\alpha}$
lead to the power-law PSD $S(f)\sim1/f^{\beta}$ with $\beta=1+\alpha/(2-\delta)$.
As a process generating the power-law probability distribution function
for $\vartheta_{j}$ a multiplicative stochastic process
\begin{equation}
\vartheta_{j+1}=\vartheta_{j}+\gamma\vartheta_{j}^{2\mu-1}+
\sigma\vartheta_{j}^{\mu}\varepsilon_{j}\label{eq:multiplicative}
\end{equation}
has been suggested. Here $\varepsilon_{k}$ are normally distributed uncorrelated
random variables with a zero expectation and unit variance. For this process
$\delta=2\mu-1$ and $\alpha=2\gamma/\sigma^{2}-2\mu$. \Eref{eq:multiplicative}
has been used for modeling of the internote interval sequences of the musical
rhythms \cite{Levitin2012}.

\section{Time-subordinated Langevin equations}

\label{sec:subord}In this section we generalize the model presented in the
previous section. We do this by noticing that in the pulse sequence there are
two strictly increasing sequences of numbers: the physical time $t$ and the
pulse number $k$. The pulse number can be interpreted as an internal time of the
pulse sequence. The relation between the physical time and the internal time is
not deterministic because the inter-pulse durations are random. Thus we propose
the introduction of the difference between the physical and the internal,
operational, time as a way to obtain $1/f$ noise also for other stochastic
processes. To do this we start with a stochastic process and interpret the time
as an internal parameter. In addition to this stochastic process we need to
include an additional relation between the physical time and the internal time.
In order to maintain a similarity to the point process described in the previous
section, the increments of the physical time should be a power-law function of
the magnitude of the signal. In this section as an initial stochastic process we
take a process described by a stochastic differential equation.

Langevin equation coupled to an additional equation for the physical time have
been introduced to describe anomalous diffusion \cite{Fogedby1994,Baule2005}.
In particular, position-dependent time subordinator has been investigated in
\cite{Srokowski2014}.

Let us start with the Langevin equation describing the diffusion of the particle
subjected to an external force
\begin{equation}
\rmd x_{t}=a(x_{t})\rmd t+b(x_{t})\rmd W_{t}\,.\label{eq:sde-initial}
\end{equation}
Here $a(x)$ and $b(x)$ are drift and diffusion coefficients and $W_{t}$ is a
standard Wiener process. For generality we assume that both coefficients $a$ and
$b$ can depend on the stochastic variable $x$. In case when the diffusion
coefficient $b$ in equation \eref{eq:sde-initial} depends on $x$ we assume It\^o
interpretation. In equation \eref{eq:sde-initial} we replace the physical time
$t$ by the operational time $\tau$,
\begin{equation}
\rmd x_{\tau}=a(x_{\tau})\rmd\tau+b(x_{\tau})\rmd W_{\tau}\,.\label{eq:x-internal}
\end{equation}
The PDF $P_{x}(x;\tau)$ of the stochastic variable $x$ as a function
of the operational time $\tau$ obeys the Fokker-Planck equation corresponding
to It\^o SDE~\eref{eq:x-internal} \cite{Gardiner2004}
\begin{equation}
\frac{\partial}{\partial\tau}P_{x}(x;\tau)=-\frac{\partial}{\partial x}a(x)P_{x}(x;\tau)
+\frac{1}{2}\frac{\partial^{2}}{\partial x^{2}}b^{2}(x)P_{x}(x;\tau)\,.
\label{eq:FP-x}
\end{equation}

Primarily we consider the situation when the small increments of the physical
time are deterministic and are proportional to the increments of the operational
time. Thus, the physical time $t$ is related to the operational time $\tau$ via
the equation
\begin{equation}
\rmd t_{\tau}=g(x_{\tau})\rmd\tau\,.\label{eq:internal-external}
\end{equation}
Here the positive function $g(x)$ is the intensity of random time that depends
on the intensity of the signal $x$. If we interpret equation
\eref{eq:sde-initial} as describing the diffusion of a particle in
non-homogeneous medium, the function $g(x)$ models the position of structures
responsible for either trapping or accelerating the particle
\cite{Srokowski2014}. Large values of $g(x)$ corresponds to trapping of the
particle, whereas small $g(x)$ leads to the acceleration of diffusion. For fixed
particle position $x$ the coefficient $g(x)$ in equation
\eref{eq:internal-external} is constant and from equation
\eref{eq:internal-external} follows the relationship
\begin{equation}
\frac{\partial}{\partial\tau}P(t;\tau|x)=-\frac{\partial}{\partial t}g(x)P(t;\tau|x)\,.
\label{eq:prob-t-tau}
\end{equation}
for the PDF $P(t;\tau|x)$ of the physical time $t$ as a function of the
operational time $\tau$. Equations \eref{eq:x-internal} and
\eref{eq:internal-external} together define the subordinated process. However,
now the processes $x(\tau)$ and $t(\tau)$ are not independent. 

Let us derive the Langevin equation for the stochastic variable $x$ in the
physical time $t$. To do this, we consider the joint PDF $P_{x,t}(x,t;\tau)$ of
the stochastic variables $x$ and $t$. Equations \eref{eq:x-internal} and
\eref{eq:internal-external} yield the two-dimensional Fokker-Planck equation
\begin{equation}
\frac{\partial}{\partial\tau}P_{x,t}(x,t;\tau)=-\frac{\partial}{\partial x}a(x)P_{x,t}
+\frac{1}{2}\frac{\partial^{2}}{\partial x^{2}}b^{2}(x)P_{x,t}
-\frac{\partial}{\partial t}g(x)P_{x,t}\,.
\label{eq:FP-x-t-tau}
\end{equation}
This equation is a combination of equations \eref{eq:FP-x} and
\eref{eq:prob-t-tau}. The zero of the physical time $t$ coincides with the zero
of the operational time $\tau$, therefore, the initial condition for equation
\eref{eq:FP-x-t-tau} is $P_{x,t}(x,t;0)=P_{x}(x,0)\delta(t)$. Coinciding zeros
of $t$ and $\tau$ lead also to the boundary condition $P_{x,t}(x,0;\tau)=0$ for
$\tau>0$, because $t$ and $\tau$ are strictly increasing.

Instead of $x$ and $t$ we can consider $x$ and $\tau$ as stochastic variables.
The stochastic variable $t$ is related to the operational time $\tau$ via
equation \eref{eq:internal-external}, therefore, the joint PDF
$P_{x,\tau}(x,\tau;t)$ of the stochastic variables $x$ and $\tau$ is related to
the PDF $P_{x,t}(x,t;\tau)$ according to the equation
\begin{equation}
P_{x,\tau}(x,\tau;t)=g(x)P_{x,t}(x,t;\tau)\,.\label{eq:transform}
\end{equation}
This equation can be obtained by noticing that the last term in equation
\eref{eq:FP-x-t-tau} contains derivative $\frac{\partial}{\partial t}$ and thus
should be equal to $-\frac{\partial}{\partial t}P_{x,\tau}$. Using equations
\eref{eq:FP-x-t-tau} and \eref{eq:transform} we get
\begin{equation}
\fl\frac{\partial}{\partial t}P_{x,\tau}(x,\tau;t)=-\frac{\partial}{\partial x}a(x)
\frac{1}{g(x)}P_{x,\tau}+\frac{1}{2}\frac{\partial^{2}}{\partial x^{2}}b^{2}(x)
\frac{1}{g(x)}P_{x,\tau}-\frac{\partial}{\partial\tau}\frac{1}{g(x)}P_{x,\tau}\,.
\label{eq:FP-x-tau-t}
\end{equation}
The PDF $P_{x,\tau}$ has the initial condition
$P_{x,\tau}(x,\tau;0)=P_{x}(x,0)\delta(\tau)$ and the boundary condition
$P_{x,\tau}(x,0;t)=0$ for $t>0$. The PDF of the subordinated random process
$x_{t}$ is $P(x,t)=\int P_{x,\tau}(x,\tau;t)\rmd\tau$. Integrating both sides of
equation \eref{eq:FP-x-tau-t} we obtain
\begin{equation}
\frac{\partial}{\partial t}P(x,t)=-\frac{\partial}{\partial x}\frac{a(x)}{g(x)}P(x,t)
+\frac{1}{2}\frac{\partial^{2}}{\partial x^{2}}\frac{b^{2}(x)}{g(x)}P(x,t)\,.
\label{eq:FP-x-external}
\end{equation}
Thus, position-dependent trapping leads to the position-dependent
coefficients in the Fokker-Planck equation, even if the initial
SDE~\eref{eq:x-internal} has constant coefficients. \Eref{eq:FP-x-external}
corresponds to the single equation in the physical time with the multiplicative
noise, 
\begin{equation}
\rmd x_{t}=\frac{a(x_{t})}{g(x_{t})}\rmd t+\frac{b(x_{t})}{\sqrt{g(x_{t})}}\rmd W_{t}\,.
\label{eq:sde-physical}
\end{equation}
In fact, the Fokker-Planck equation \eref{eq:FP-x-tau-t} can be obtained from
the coupled equations \eref{eq:sde-physical} and 
\begin{equation}
\rmd\tau_{t}=\frac{1}{g(x_{t})}\rmd t\,.
\label{eq:external-internal}
\end{equation}

The relationship between the physical time $t$ and the operational time $\tau$
can be not necessarily deterministic, equation \eref{eq:internal-external} can
have a stochastic term. If the fluctuations of this stochastic term are much
faster than the fluctuations of the stochastic variable $x$, we can approximate
them by the average value. In this case $g(x)$ describes the average increment
of the physical time. If this average is positive, the derivation presented
above is still valid and equation \eref{eq:sde-physical} holds.

\section{Example equations generating signals with $1/f$ noise}

\label{sec:example}In this section we consider several stochastic processes and,
introducing the internal and external times, we check whether $1/f$ noise can be
obtained. In a signal consisting of pulses the internal time is just the pulse
number and the increment of the physical time is equal to the inter-pulse
duration. The intensity of such a signal is inversely proportional to the
inter-pulse duration. In order to obtain $1/f$ noise similarly as for the signal
consisting of pulses we choose the function $g(x)$ in equation
\eref{eq:internal-external} as a power-law function of $x$, $g(x)\sim
x^{-2\eta}$, where $\eta$ is the power-law exponent.

Let us start from a simple Brownian motion 
\begin{equation}
\rmd x_{\tau}=\rmd W_{\tau}\,.\label{eq:brownian}
\end{equation}
In order to keep the stochastic variable $x$ always positive we include
reflective boundary at $x=x_{\mathrm{min}}>0$. We consider equation
\eref{eq:brownian} together with the relation 
\begin{equation}
\rmd t_{\tau}=x_{\tau}^{-2\eta}\rmd\tau
\label{eq:phys-intern-2}
\end{equation}
between the physical time $t$ and internal time $\tau$. According to
\eref{eq:sde-physical} the resulting equation in the physical time is 
\begin{equation}
\rmd x_{t}=x_{t}^{\eta}\rmd W_{t}\,.
\end{equation}
More generally, the initial equation can include a position-dependent force. If
we take the equation describing the Bessel process 
\begin{equation}
\rmd x_{\tau}=\left(\eta-\frac{\lambda}{2}\right)\frac{1}{x_{\tau}}\rmd\tau
+\rmd W_{\tau}
\label{eq:bessel}
\end{equation}
together with equation \eref{eq:phys-intern-2}, then the resulting equation in
the physical time becomes 
\begin{equation}
\rmd x_{t}=\left(\eta-\frac{\lambda}{2}\right)x_{t}^{2\eta-1}\rmd t
+x_{t}^{\eta}\rmd W_{t}\,.
\label{eq:sde-1}
\end{equation}
Here the parameter $\lambda$ gives the power-law exponent of the steady-state PDF.
The same equation \eref{eq:sde-1} in physical time arises starting from the
geometric Brownian motion, 
\begin{equation}
\rmd x_{\tau}=\left(\eta-\frac{\lambda}{2}\right)x_{\tau}\rmd\tau
+x_{\tau}\rmd W_{\tau}\,,
\end{equation}
and the relation between the internal time and the physical time
\begin{equation}
\rmd t_{\tau}=x_{\tau}^{-2(\eta-1)}\rmd\tau\,.
\end{equation}

Nonlinear SDE~\eref{eq:sde-1} for generating signals with $1/f^{\beta}$ spectrum
has been proposed in \cite{Kaulakys2004,Kaulakys2006}. As has been shown in
\cite{Ruseckas2014}, the reason for the appearance of $1/f$ spectrum is the
scaling properties of the signal: the change of the magnitude of the variable
$x\rightarrow ax$ is equivalent to the change of the time scale $t\rightarrow
a^{2(\eta-1)}t$. Connection of the power-law exponent $\beta$ in the PSD with
the parameters of equation \eref{eq:sde-1} is given by the equation
\cite{Kaulakys2006,Ruseckas2014}
\begin{equation}
\beta=1+\frac{\lambda-3}{2(\eta-1)}\,.
\label{eq:beta}
\end{equation}
Analysis \cite{Ruseckas2010} of SDE~\eref{eq:sde-1} shows that equation
\eref{eq:beta} is valid only for the values of the parameters $\eta$ and
$\lambda$ yielding $0 < \beta < 2$.

Nonlinear SDE~\eref{eq:sde-1} leads to the stationary process and non-diverging
steady state PDF only when the diffusion of stochastic variable $x$ is
restricted. The simplest choice of the restriction is the reflective boundary
conditions at $x=x_{\mathrm{min}}$ and $x=x_{\mathrm{max}}$. The presence of the
restrictions of diffusion makes the scaling properties of equation
\eref{eq:sde-1} only approximate and limits the power-law part of the PSD to a
finite range of frequencies. This range of frequencies has been qualitatively
estimated as \cite{Ruseckas2014}
\begin{eqnarray}
x_{\mathrm{min}}^{2(\eta-1)} \ll 2\pi f\ll x_{\mathrm{max}}^{2(\eta-1)}\,,\qquad\eta>1\,,\\
x_{\mathrm{max}}^{-2(1-\eta)} \ll 2\pi f\ll x_{\mathrm{min}}^{-2(1-\eta)}\,,\qquad\eta<1\,.\nonumber 
\end{eqnarray}
By increasing the ratio $x_{\mathrm{max}}/x_{\mathrm{min}}$ one can get an
arbitrarily wide range of the frequencies where the PSD has $1/f^{\beta}$
behavior.

\begin{figure}
\includegraphics[width=0.45\textwidth]{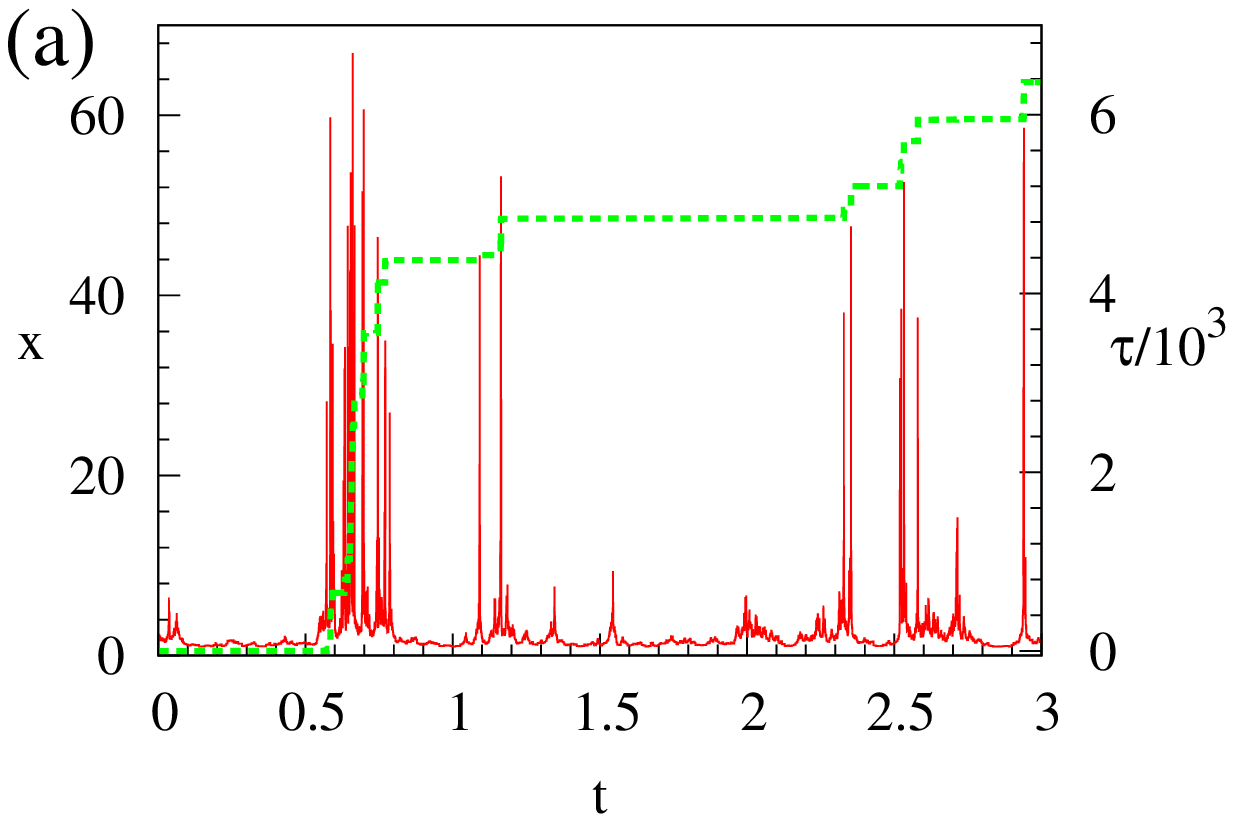}\hspace{1cm}\includegraphics[width=0.45\textwidth]{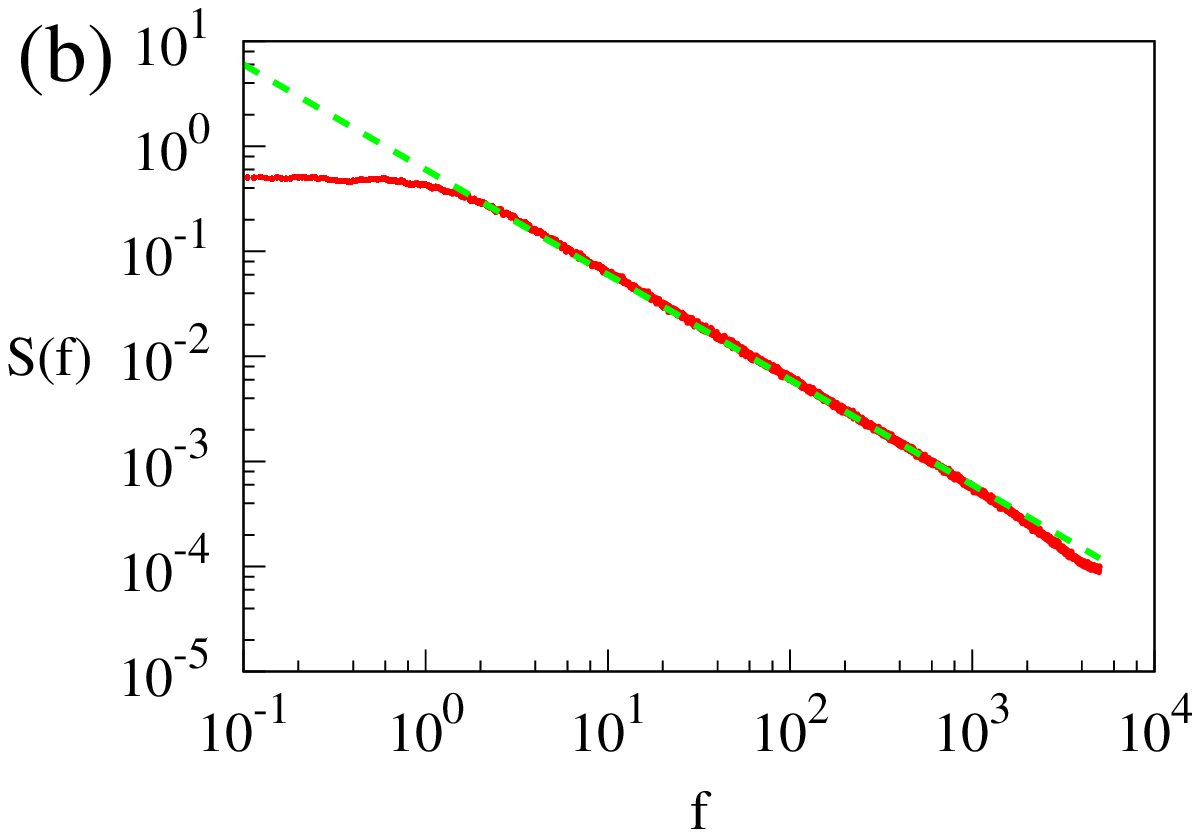}
\caption{(a) Signal generated by equation \eref{eq:sde-1} with the parameters
$\eta=5/2$ and $\lambda=3$ (solid red line) together with the corresponding
internal time (dashed green line). Reflective boundaries at $x_{\mathrm{min}}=1$
and $x_{\mathrm{max}}=1000$ have been used. (b) PSD
of the generate signal. Dashed green line shows the slope $1/f$.}
\label{fig:nonlin-sde}
\end{figure}

An example of a signal generated by equation \eref{eq:sde-1} together with the
internal time $\tau$ is shown in \fref{fig:nonlin-sde}(a). We used the
parameters $\eta=5/2$, $\lambda=3$ and reflective boundaries at
$x_{\mathrm{min}}=1$ and $x_{\mathrm{max}}=1000$. The method of numerical
solution is discussed in the next section. We see that internal time $\tau$
increases rapidly when the signal $x$ acquires large values and $\tau$ changes
slowly when $x$ is small. According to equation \eref{eq:beta} the choice of
$\lambda=3$ should result in $1/f$ behavior of the PSD. The corresponding power
spectral density $S(f)$ is shown in \fref{fig:nonlin-sde}(b). The numerical
solution of the equation confirms a presence of a wide region of frequencies
where the spectrum has $1/f$ behavior.

When the stochastic variable $x$ can acquire both positive and negative values,
the function $g(x)$ cannot be just a simple power-law, because $g(x)$ becomes
unbounded or equal to zero when $x\rightarrow0$. In order to avoid this problem
we require that function $g(x)$ should have power-law behavior only
asymptotically, for large vales of $|x|$. One of the possible choices is
\begin{equation}
g(x)=\frac{1}{(x^{2}+x_{0}^{2})^{\eta}}\,.
\end{equation}
Here we added a constant $x_{0}$ that corrects the behavior of the function
$g(x)$ at $x=0$. The power-law behavior is preserved when $|x|\gg x_{0}$.

The stochastic variable $x$ can acquire both positive and negative values if we
start from the Ornstein-Uhlenbeck process 
\begin{equation}
\rmd x_{\tau}=-\gamma x_{\tau}\rmd\tau+\rmd W_{\tau}\,.
\label{eq:orn-uhl}
\end{equation}
Here the parameter $\gamma$ is the relaxation rate. We consider equation
\eref{eq:orn-uhl} together with the relation 
\begin{equation}
\rmd t_{\tau}=\frac{1}{(x_{\tau}^{2}+x_{0}^{2})^{\eta}}\rmd\tau
\label{eq:phys-intern-3}
\end{equation}
between the physical time $t$ and internal time $\tau$. According to
\eref{eq:sde-physical}, equations \eref{eq:orn-uhl} and \eref{eq:phys-intern-3}
leads to SDE 
\begin{equation}
\rmd x_{t}=-\gamma(x_{t}^{2}+x_{0}^{2})^{\eta}x_{t}\rmd t
+(x_{t}^{2}+x_{0}^{2})^{\frac{\eta}{2}}\rmd W_{t}
\label{eq:sde-4}
\end{equation}
in the physical time $t$. \Eref{eq:sde-4} can be written as
\begin{equation}
\rmd x_{t}=\left(-\frac{x_{t}^{2}+x_{0}^{2}}{x_{\mathrm{max}}^{2}}\right)
(x_{t}^{2}+x_{0}^{2})^{\eta-1}x_{t}\rmd t+(x_{t}^{2}+x_{0}^{2})^{\frac{\eta}{2}}\rmd W_{t}
\end{equation}
where
\begin{equation}
x_{\mathrm{max}}=\frac{1}{\sqrt{\gamma}}
\end{equation}
defines a cut-off position at large values of $x$.

Another interesting equation describing the evolution in internal time is 
\begin{equation}
\rmd x_{\tau}=\left(\eta-\frac{\lambda}{2}\right)\frac{x_{\tau}}{x_{\tau}^{2}+x_{0}^{2}}
\rmd\tau +\rmd W_{\tau}\,.
\label{eq:sde-3}
\end{equation}
In this equation the relaxation rate depends on the magnitude of the signal. If
$|x|\ll x_{0}$ we get the equation of Ornstein-Uhlenbeck type, whereas for
large values of $|x|$ the relaxation decreases with increasing $|x|$.
\Eref{eq:sde-3} together with \eref{eq:phys-intern-3} result in the following
equation in the physical time:
\begin{equation}
\rmd x_{t}=\left(\eta-\frac{\lambda}{2}\right)(x_{t}^{2}+x_{0}^{2})^{\eta-1}x_{t}
\rmd t+(x_{t}^{2}+x_{0}^{2})^{\frac{\eta}{2}}\rmd W_{t}\,.
\label{eq:sde-q-gauss}
\end{equation}
Finally, the combination of equations \eref{eq:orn-uhl} and \eref{eq:sde-3}, 
\begin{equation}
\rmd x_{\tau}=-\left(\gamma-\left(\eta-\frac{\nu}{2}\right)\frac{1}{x_{\tau}^{2}
+x_{0}^{2}}\right)x_{\tau}\rmd\tau+\rmd W_{\tau}\,, 
\end{equation}
together with \eref{eq:phys-intern-3} leads to a more general equation
in the physical time
\begin{equation}
\rmd x_{t}=\left(\eta-\frac{\nu}{2}-\frac{x_{t}^{2}+x_{0}^{2}}{x_{\mathrm{max}}^{2}}\right)
(x_{t}^{2}+x_{0}^{2})^{\eta-1}x_{t}\rmd t
+(x_{t}^{2}+x_{0}^{2})^{\frac{\eta}{2}}\rmd W_{t}\,.
\end{equation}
Nonlinear SDE~\eref{eq:sde-q-gauss} has been investigated in
\cite{Ruseckas2011}. It has been shown that SDE~\eref{eq:sde-q-gauss} generates
a signal with the steady-state PDF described by the $q$-Gaussian distribution
featuring in the non-extensive statistical mechanics. In addition, the spectrum
of the generated signal has $1/f^{\beta}$ behavior in a wide range of
frequencies, with the power-law exponent $\beta$ given by equation
\eref{eq:beta}.

\begin{figure}
\includegraphics[width=0.45\textwidth]{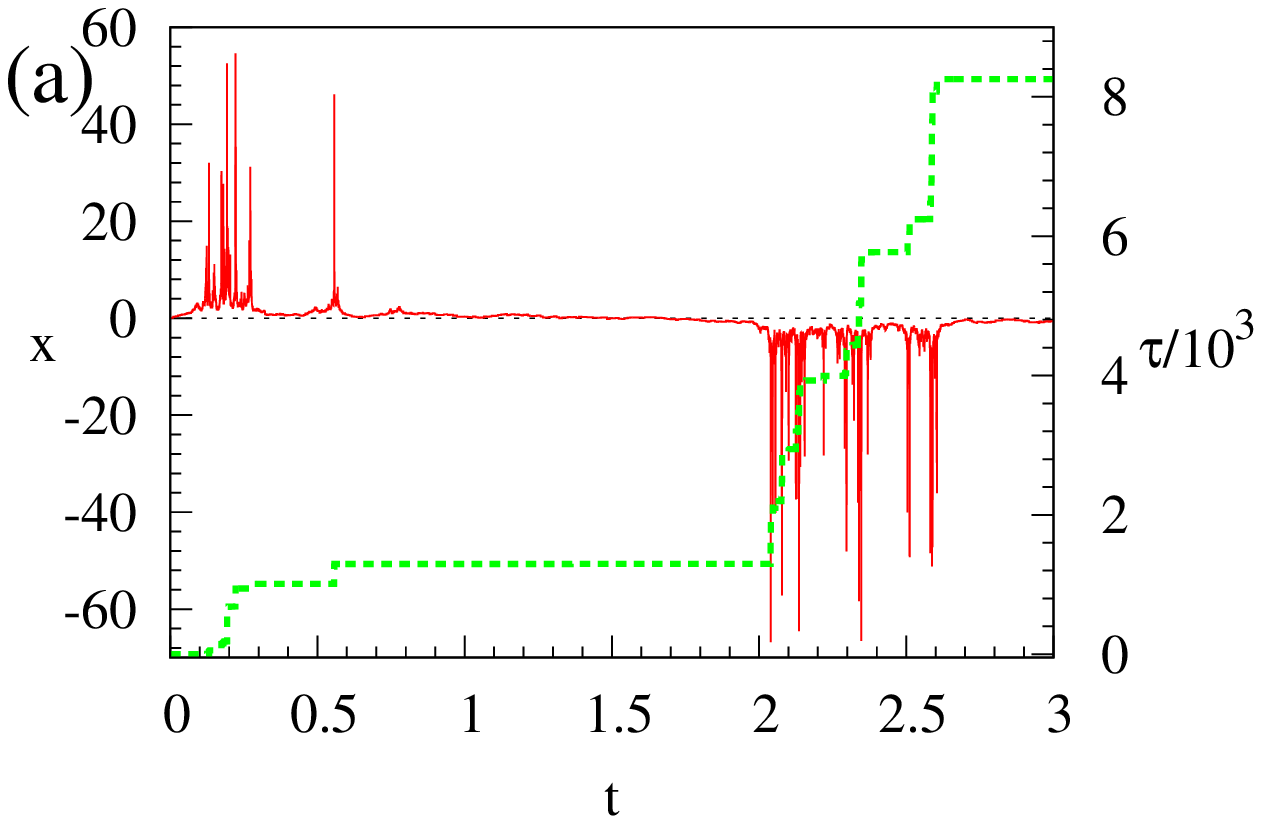}\hspace{1cm}\includegraphics[width=0.45\textwidth]{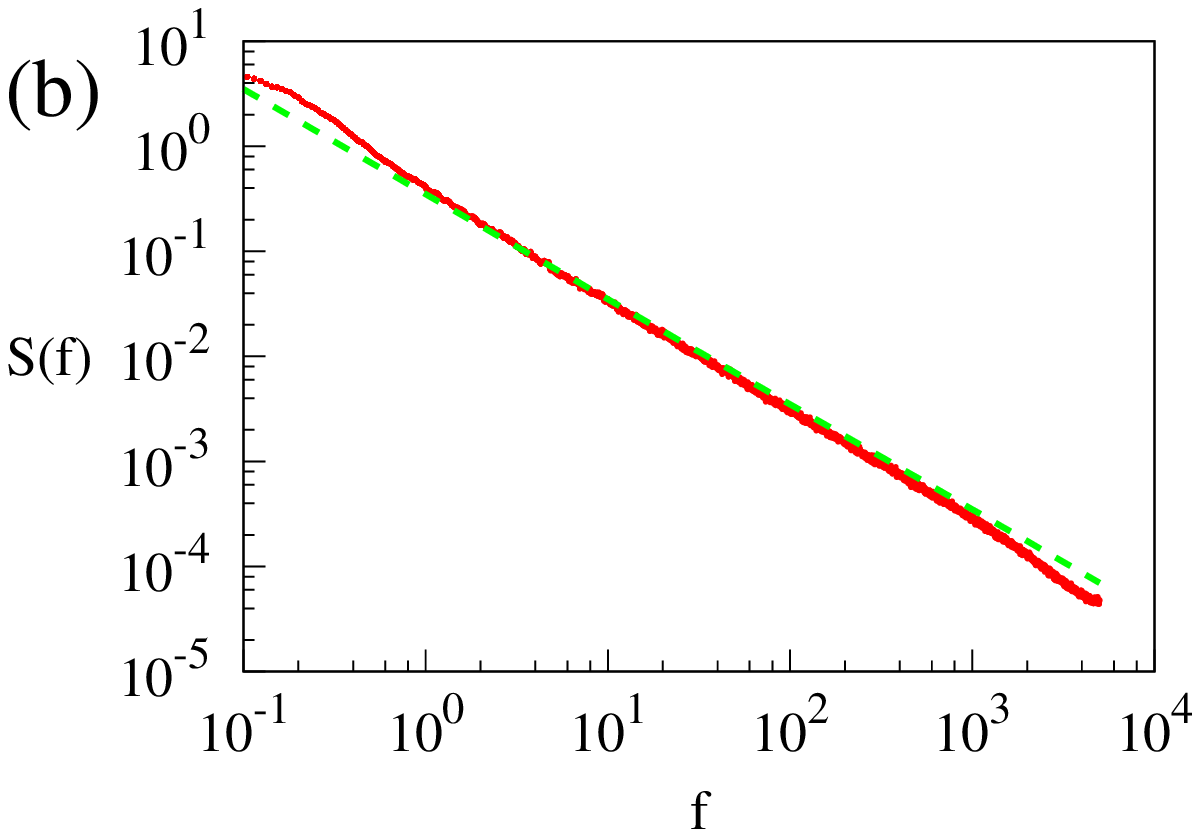}
\caption{(a) Signal generated by equation \eref{eq:sde-q-gauss} with the
parameters $\eta=5/2$, $\lambda=3$ and $x_{0}=1$ (solid red line) together with
the corresponding internal time (dashed green line). (b) PSD 
of the generate signal. Dashed green line shows the slope $1/f$.}
\label{fig:q-gaussian}
\end{figure}

An example of a signal generated by equation \eref{eq:sde-q-gauss} together with
the internal time is shown in \fref{fig:q-gaussian}(a). We used the parameters
$\eta=5/2$, $\lambda=3$ and $x_{0}=1$. We see that the internal time $\tau$
increases rapidly when the absolute value of the signal $x$ is large and $\tau$
changes slowly when the absolute value of $x$ is small. The internal time $\tau$
increases both for positive and negative values of $x$. 
The PSD of a signal generated by equation \eref{eq:sde-q-gauss} is shown in
\fref{fig:q-gaussian}(b). The numerical solution confirms a presence of a region
where the spectrum behaves as $1/f$. Thus the introduction of negative values of
$x$ does not destroy $1/f$ spectrum. 

\section{Numerical approach}

\label{sec:numer}Introduction of the internal time can be an effective technique
for solution of highly non-linear SDEs. For numerical solution of nonlinear
equations the solution schemes involving a fixed time step $\Delta t$ can be
inefficient. For example, in equation \eref{eq:sde-1} with $\eta>1$ large values
of stochastic variable $x$ lead to large coefficients and thus require a very
small time step. The numerical solution scheme can by improved by introducing
the internal time $\tau$ that is different from the real, physical, time $t$.

Let us consider equation \eref{eq:sde-1} with the noise multiplicativity
exponent $\eta>1$. We can introduce internal time $\tau$ using the equation
\begin{equation}
\rmd\tau_{t}=x_{t}^{2\eta}\rmd t\,.
\end{equation}
Then, according to equations \eref{eq:sde-physical} and
\eref{eq:external-internal}, SDE~\eref{eq:sde-1} is equivalent to coupled
equations
\begin{eqnarray}
\rmd x_{\tau} = \left(\eta-\frac{\nu}{2}\right)\frac{1}{x_{\tau}}\rmd\tau+\rmd W_{\tau}\,,
\label{eq:sde-5}\\
\rmd t_{\tau} = \frac{1}{x_{\tau}^{2\eta}}\rmd\tau\,.
\end{eqnarray}
Now, equation \eref{eq:sde-5} is much simpler than the initial equation
\eref{eq:sde-1}. Discretizing the internal time $\tau$ with the step
$\Delta\tau$ and using the Euler-Marujama approximation for the
SDE~\eref{eq:sde-5} we get
\begin{eqnarray}
x_{k+1}= x_{k}+\left(\eta-\frac{\lambda}{2}\right)\frac{1}{x_{k}}\Delta\tau
+\sqrt{\Delta\tau}\varepsilon_{k}\,,\label{eq:discr-1}\\
t_{k+1}= t_{k}+\frac{\Delta\tau}{x_{k}^{2\eta}}\,.\label{eq:discr-2}
\end{eqnarray}
Here $\varepsilon_{k}$ are normally distributed uncorrelated random variables.
Equations \eref{eq:discr-1} and \eref{eq:discr-2} provide the numerical method
for solving equation \eref{eq:sde-1}. One can interpret equations
\eref{eq:discr-1}, \eref{eq:discr-2} as an Euler-Marujama scheme with a variable
time step $\Delta t_{k}=\Delta\tau/x_{k}^{2\eta}$ that adapts to the
coefficients in the equation. The cost of the introduction of the internal time
is the randomness of the increments of the real, physical time $t$. To get the
discretization of time with fixed steps the signal generated in such a way
should be interpolated.

Another possible choice is to introduce the internal time $\tau$ by the equation
\begin{equation}
\rmd\tau_{t}=x_{t}^{2(\eta-1)}\rmd t\,.
\end{equation}
In this case we obtain a different pair of equations
\begin{eqnarray}
\rmd x_{\tau} = \left(\eta-\frac{\lambda}{2}\right)x_{\tau}\rmd\tau+x_{\tau}\rmd W_{\tau}\,,
\label{eq:sde-6}\\
\rmd t_{\tau} = \frac{1}{x_{\tau}^{2(\eta-1)}}\rmd\tau\,.
\end{eqnarray}
Note, that now the internal time $\tau$ is dimensionless even if $x$ and $t$ are
not. Discretizing the internal time $\tau$ with the step $\Delta\tau$ and using
the Euler-Marujama approximation for the SDE~\eref{eq:sde-6} we obtain
\begin{eqnarray}
x_{k+1} = x_{k}+\left(\eta-\frac{\lambda}{2}\right)x_{k}\Delta\tau
+x_{k}\sqrt{\Delta\tau}\varepsilon_{k}\,.\\
t_{k+1} = t_{k}+\frac{\Delta\tau}{x_{k}^{2(\eta-1)}}\,.
\end{eqnarray}
This method of solution has been proposed in \cite{Kaulakys2004}. On the other
hand, using Milstein approximation for the SDE~\eref{eq:sde-6} we have
\begin{eqnarray}
x_{k+1} = x_{k}+\left(\eta-\frac{\lambda}{2}\right)x_{k}\Delta\tau
+x_{k}\sqrt{\Delta\tau}\varepsilon_{k}+\frac{1}{2}x_{k}\Delta\tau(\varepsilon_{k}^{2}-1)\,,
\label{eq:discr-3}\\
t_{k+1} = t_{k}+\frac{\Delta\tau}{x_{k}^{2(\eta-1)}}\,.
\end{eqnarray}
Note, that the last term in equation \eref{eq:discr-3} differs from the
corresponding term in the equation obtained just by using a variable time step
$\Delta t=\Delta\tau/x_{k}^{2(\eta-1)}$ in the Milstein approximation for
equation \eref{eq:sde-1}.

Numerical simulation of subordinated equations using fixed step of operational
time and random increment of physical time has been discussed in
\cite{Kleinhans2007,Magdziarz2007}. Variable time step makes numerical
simulation in \cite{Kleinhans2007,Magdziarz2007} similar to the method proposed
in this section. The main difference of our method from previous discussions
of subordinated equations lies in the depence of the increment of the physical
time on the magnitude of the signal $x$.

\section{Discussion and conclusions}

\label{sec:concl}
In summary, we have demonstrated that starting from a random process described
by a SDE and introducing the difference between the internal time and the
physical time $1/f$ behavior of the PSD can be obtained.

One of the physical situations where the difference between the internal and
physical time can arise is transport in an inhomogeneous medium. Impurities and
regular structures in the medium can cause transport of variable speed, the
particle may be trapped for some time or accelerated. Nonhomogeneous systems
exhibit not only subdiffusion related to traps, but also enhanced diffusion as a
result of the disorder. For example, movement of particles between two
neighboring lattice sites in an interacting particle system is superdiffusive
due to the disorder and subdiffusive without the disorder \cite{Ben-Naim2009}.
The dynamics in a medium with traps is described by the continuous time random
walk theory (CTRW) \cite{Metzler2000,Metzler2004}. In a description equivalent to
CTRW the dynamics of the particle is Markovian and governed by the Langevin
equation in an auxiliary, operational, time instead of the physical time. This
Markovian process is subordinated to the process yielding the physical time.

In the case of subdiffusion the PSD of the signals generated by subordinated
Langevin equations has power-law behavior $S(f)\sim f^{\alpha - 1}$ as
$f\rightarrow 0$ \cite{Yim2006}, where $\alpha$ is the power-law exponent in the
time dependence of the mean square displacement. Since for subdiffusion $0 <
\alpha < 1$, the power-law exponent $\beta$ in the PSD is smaller than $1$. The results
obtained in this paper suggest that $1/f$ noise in subdiffusion should occur in
heterogeneous medium, where trapping time depends on the position
\cite{Kazakevicius2015}.

The traditional CTRW provides a homogeneous description of the medium. More
complex situation is the diffusion in nonhomogeneous media, for example, 
diffusion on fractals and multifractals \cite{Schertzer2001}. Heterogeneous
medium with steep gradients of the diffusivity can be created via a local
variation of the temperature in thermophoresis experiments
\cite{Maeda2012,Mast2013}. Spatial heterogeneities are also present in the case
of anomalous diffusion in subsurface hydrology \cite{Dentz2010}. In the random
walk description spatially varying diffusivity can be translated into a local
dependence of the waiting time for a jump event. In the heterogeneous medium the
properties of a trap can reflect the medium structure, thus in the description
of transport the waiting time should explicitly depend on the position of the
particle \cite{Srokowski2014}. A method to include position dependent waiting
time is a consideration of the position-dependent time subordinator
\cite{Srokowski2014}.

In general, the trapping time can depend not on the position of the particle but
on some other quantity. Then in the dynamics of this quantity the difference
between the physical and operational time also arises, with the relationship
between the times dependent on the intensity of the signal.

In socio-economical systems the internal time can reflect fluctuating human
activity \cite{Mathiesen2013}. For example, in finance the long-range
correlations in volatility arise due to fluctuations of the trading activity
\cite{Plerou2001,Gabaix2003}.

We have shown that $1/f$ noise occurs when the internal time and the physical
time are related via the power-law function of the signal intensity, for
example, via equation \eref{eq:phys-intern-2} or \eref{eq:phys-intern-3}.
Although we have considered only random processes described by a SDE, we expect
that the mechanism of the appearance of $1/f$ noise presented here is quite
general and should work also for other random processes. We anticipate that the
present model can be useful for explaining $1/f$ noise in different complex
systems.

In addition, we suggested a way of solving highly non-linear SDEs by introducing
suitably chosen internal time and variable step of integration.

\section*{References}

\providecommand{\newblock}{}

\end{document}